\documentclass{emulateapj}

\usepackage{graphicx}

\begin{document}

\title{FEEDBACK FROM CENTRAL BLACK HOLES IN ELLIPTICAL GALAXIES:
  TWO-DIMENSIONAL MODELS COMPARED TO ONE-DIMENSIONAL MODELS}

\author
{Gregory S. Novak,\altaffilmark{1}
Jeremiah P. Ostriker,\altaffilmark{1}, and Luca
Ciotti\altaffilmark{2}}

\altaffiltext{1}{Department of Astrophysical Sciences, Peyton Hall, Princeton
University, Princeton, NJ 08544, USA}
\altaffiltext{2}{Department of Astronomy, University of Bologna, via Ranzani 1,
I-40127, Bologna, Italy}

\shorttitle{TWO-DIMENSIONAL AGN SIMULATIONS}
\shortauthors{NOVAK, OSTRIKER, \& CIOTTI}


\begin{abstract}  
  We extend the black hole (BH) feedback models of Ciotti, Ostriker, and
  Proga to two dimensions.  In this paper, we focus on identifying the
  differences between the one-dimensional and two-dimensional
  hydrodynamical simulations.  We examine a normal, isolated $L_*$
  galaxy subject to the cooling flow instability of gas in the inner
  regions.  Allowance is made for subsequent star formation, Type Ia
  and Type II supernovae, radiation pressure, and inflow to the
  central BH from mildly rotating galactic gas which is being
  replenished as a normal consequence of stellar evolution.  The
  central BH accretes some of the infalling gas and expels a
  conical wind with mass, momentum, and energy flux derived from both
  observational and theoretical studies.  The galaxy is assumed to
  have low specific angular momentum in analogy with the existing
  one-dimensional case in order to isolate the effect of
  dimensionality.  The code then tracks the interaction of the
  outflowing radiation and winds with the galactic gas and their
  effects on regulating the accretion.  After matching physical
  modeling to the extent possible between the one-dimensional and two-dimensional treatments, we
  find essentially similar results in terms of BH growth and
  duty cycle (fraction of the time above a given fraction of the
  Eddington luminosity).  In the two-dimensional calculations, the cool shells
  forming at 0.1--1 kpc from the center are Rayleigh--Taylor unstable
  to fragmentation, leading to a somewhat higher accretion rate, less
  effective feedback, and a more irregular pattern of bursting
  compared to the one-dimensional case.
\end{abstract}


\maketitle

\section{Introduction}
Nearly all massive galaxies are thought to harbor supermassive black
holes (SMBHs), and the properties of the black holes (BHs) are known to be
well correlated with those of their host galaxies \citep{gebhardt:00,
  ferrarese:00, tremaine:02, novak:06-blackholes, gueltekin:09}.  The
direction of the causal link between SMBHs and host galaxy properties
remains unclear, but the existence of such a correlation suggests an
intimate connection between the physics of BH growth and galaxy
formation.  There has been intense theoretical interest in the ability
of active galactic nuclei (AGNs) to affect the star formation histories
and observed colors of galaxies \citep[e.g.,][]{dimatteo:05, croton:06}.
There is certainly sufficient energy and momentum available to have a
profound effect on the structure and dynamics of the interstellar
medium (ISM).

Many physical processes can act to couple accretion by the central BH
to the galaxy's ISM.  The most obvious and well-observed result of
SMBH accretion is the prodigious luminous output.  Radiatively
efficient accretion converts a significant fraction of the rest mass
energy of the infalling gas to radiation, and the bulk of the matter
that enters BHs over the history of the universe is known to
do so in a radiatively efficient manner \citep{soltan:82}.  The
emitted photons impart energy and momentum to the galaxy's ISM via
electron scattering, photoionization, scattering due to atomic
resonance lines, and absorption by dust grains.

Accreting SMBHs are also known to drive broad absorption line (BAL)
winds that convey mass, momentum, and energy to the surrounding
galaxy.  These winds are launched by electromagnetic processes within
a few hundred gravitational radii of the central BH
\citep[e.g.,][]{proga:00}, but by the time the wind galactic length
scales, the energy is carried by the kinetic motion of the gas.  Only
fraction of the energy output of the AGN will be converted to a wind,
that wind will proceed to inject nearly all of its energy and momentum
into the galaxy's ISM.  This high interaction efficiency has inspired
strong theoretical interest in the effect of mechanical AGN feedback
on galaxy properties \citep{dimatteo:05, springel:05-blackholes,
  johansson:09}.

The combined effect of these feedback processes can dramatically
affect the galaxy and act to reduce subsequent SMBH accretion.
Physical modeling of the complex processes involved is beginning to
reach a level where observational tests of the models are possible.
Building on previous work in one dimension \citep{ciotti:97,
  ciotti:01, ciotti:07, ciotti:09-feedback, shin:10, ciotti:10}, we
perform two-dimensional simulations of SMBH accretion in the context
of normal but isolated elliptical galaxies.  Thus, the source of
accreted gas is the fraction ($\sim M_*/6$) ejected into the
ISM due to normal processed of stellar evolution,
rather than the comparable amounts of gas added via cosmological
infall.  We take care to resolve the inner length scales of accretion
(the Bondi radius) while also running the simulations long enough to
resolve galactic length and timescales.  Many physical processes must
be included and numerically a large dynamic range is needed.
Although quite well developed from the point of view of the included
physics, our previous one-dimensional simulations obviously could not address the
impact of genuinely two-dimensional phenomena on the feedback
phenomenon. We recall here the most important of them, that will be
discussed in this paper.

The first one concerns the possible instabilities
(e.g., Rayleigh--Taylor, RT) that can affect the evolution of the cold
shells that appear during the evolution of one-dimensional models.  The central
bursts are caused by these shells, and the shells have a two-fold
origin.  At the beginning of each major burst, the classical Field
cooling instability \citep{field:65} appears around 1 kpc from the center, due to the
local critical balance between heating and cooling. As the density increases,
the shell starts to fall toward the center and compresses the
gas. When a burst first appears, shock waves are sent from the center toward the
falling shell, and a series of sub-bursts and consequent reflected
shock waves impact on the cold shell, increasing its density still
further.  As the bulk of star formation in the one-dimensional models happens in
these cold shells, it is very important to understand the cold shell
physics, not only from the point of view of central accretion, but
also for the starburst which occurs within the cold
shell. From the short description above, a few obvious questions
arise: for example, will the allowance of the additional degree of
freedom still lead to a formation of a cold shell near the center in
the case of an aspherical galaxy? What is the effect of non-zero angular
momentum in the gas?  Will the cold shell fall toward the
center as in the one-dimensional simulations or it will break up due to
(RT)
instability? Even more important, what is the fate of the 
multiple interacting shocks?  Will the accretion still be characterized by
strong bursts separated by long time intervals or will the breakup of
the shells lead to cold fingers of dense gas being accreted in a more
or less steady flow while hot gas flows outward, therefore resulting
in flows that at each radius are partially accreting and partially
outflowing?

The second reason to move to two-dimensional simulations is to explore the
interaction (and the consequent mechanical feedback) of the conical
nuclear wind with the galaxy ISM. In our previous one-dimensional simulations
this interaction was necessarily described as a spherical average of
an inherently non-spherical effect, even
though we had taken into account several physical aspects of the
phenomenon via a time-dependent differential equation.  Clearly, a two-dimensional
simulation is also needed to explore Kelvin--Helmholtz instabilities at the
interface between the outflowing conical wind and the ISM.

In the present work, we focus on two-dimensional simulations of a galaxy with very
low specific angular momentum in order to make close contact with the
existing one-dimensional simulations.  We would like to isolate the effect of
increasing the dimensionality of the simulation.  The adopted angular
momentum profile is consistent with the slowest of the SAURON
slow-rotators \citep{emsellem:04}.  In future work, we will expand our
treatment to include angular momentum transport via the standard
$\alpha$ prescription \citep{shakura:73} and more recent models based
on gravitational torques \citep{hopkins:10-analytic-preprint}.

There have been many numerical simulations of SMBH accretion and the
subsequent effects on the galaxies containing the resulting AGN.
Nearly all of the efforts to date can be classified into three broad categories.
\citet{dimatteo:05}, \citet{debuhr:10,debuhr:11}, and
\citet{johansson:09} are examples where the simulations cover length
scales from $\simeq$ 100 pc to tens of kpc and timescales from a
fraction of a Myr to several Gyr.  Galactic length and timescales are
resolved, but the SMBH accretion and feedback processes are considered
to be sub-resolution.  Complementary studies by \citet{kurosawa:09-2d}
and \citet{kurosawa:09-3d} are examples of multi-dimensional
simulations that cover the length scales from a few AU to $\simeq$ 1
pc.  Length and timescales relevant to SMBH accretion are resolved,
and the generation of radiatively driven winds is computed, but these
simulations do not approach galactic length or timescales, and infall
rates are taken as given.  \citet{hopkins:10-simulation} and
\citet{levine:08} are examples of a multi-resolution studies of SMBH
accretion involving progressively higher spatial resolution
simulations run for progressively shorter times.  The
highest spatial resolution simulations go down to a fraction of a pc
and are run for about one Myr of simulation time.  These simulations
spatially resolve the accretion process, but do not reach galactic
timescales.  Therefore, they cannot self-consistently calculate the
effect of AGN feedback on the gas in the galaxy as a whole and the
subsequent SMBH accretion.  

Finally, there have been several numerical studies of accretion by
intermediate-mass black holes (IMBHs) with the goal of understanding
BH growth in the early universe \citep{alvarez:09, park:10-preprint}.
In terms of dimensionless length scales $r/r_{\rm Bondi}$ or $r/r_{\rm
  Schwarzschild}$, some of these simulations are similar to our
simulations.  However, studies of IMBH accretion focus on BHs
with very small masses (100--1000 $M_\odot$) compared to those
presented here.  Therefore, the relevant physical length and
timescales are much smaller and the physical sources of the infalling gas
are very different from those considered in the present work.

Our goal is to resolve both the relevant accretion length and
timescales while at the same time resolving galactic length and
timescales \citep[cf.][]{levine:08, alvarez:09}.  
There have been only a few attempts to perform multi-dimensional
simulations that bridge the gap between galactic and SMBH scales,
although several papers have examined the interaction
between an outflowing wind/jet with specified properties and the surrounding
intergalactic medium
\citep[cf.][]{metzler:94,omma:04,sijacki:07,sternberg:08,reeves:09,fabian:09,arieli:10}

The present work is an attempt to simultaneously resolve the inner
length scales relevant to SMBH accretion (a few pc), outer length scales
relevant to galaxies (tens of kpc), inner timescales relevant to SMBH
accretion (a few years), and outer timescales relevant to galaxies and
stellar evolution (10 Gyr).  However, the region inside of 1 pc
including the disk and the SMBH itself are still treated as sub-resolution
physics and we compute the output from these regions as time-dependent
functions of the input to them, utilizing formulae from the above
quoted sources.

We take particular care to resolve the inner scales where the rate of
accretion is set (the Bondi radius) even for the hot gas in the
system.  This is important because heating due to AGN radiation or mechanical energy input may raise the gas
temperature as it falls toward the SMBH, decreasing the relevant Bondi
radius.  For example, \citet{alvarez:09} performed cosmological simulations of
accretion onto the SMBH left by the first stars at high redshift.  Their
resolution was sufficient to resolve the Bondi radius for the cold gas
in their simulation, but not for the much hotter gas affected by
feedback.  This means that if the heating would have been effective
between the resolution of their simulation and the actual hot-gas
Bondi radius, then in reality the gas would never have made it to the SMBH.
It is important to realize that the material emitted by the evolving
stars (the fuel for SMBH accretion) is thermalized by stellar
velocity dispersion to the equivalent temperature, and that the
stellar velocity dispersion near the BH increases (as $1/r$ in the
isotropic case). Therefore, the gas is injected at higher and higher
temperature approaching the central BH and the location of the Bondi
radius does not change significantly even if the SMBH grows in mass.
Care must be exercised in the numerical treatment of the central
regions. To put it simply, Bondi-type accretion cannot be computed properly
unless the detailed thermal state of the gas is followed to well
within the radius at which gravitational and thermal energies are
balanced.

Section \ref{sec:methods} describes the simulations and our basic
approach, Section \ref{sec:results} gives our results, and
Section \ref{sec:conclusions} summarizes our conclusions.

\section{Methods}
\label{sec:methods}
We use the Zeus hydrodynamics code \citep{stone:92-hd} because of its
ability to utilize spherical geometry and variable cell sizes.  The
code is not adaptive in the sense that it does not dynamically decide
where to place extra cells, but the variable cell sizes allow us to
identify a center of interest and take advantage of increased
resolution near that point.  We extend the code with mass, energy, and
momentum source terms appropriate for stellar evolution in elliptical
galaxies, SMBH accretion, and feedback resulting from the accretion,
following the treatment described in \citet{ciotti:09-feedback}.

The simulation grid is the meridional plane in spherical coordinates
where all quantities are assumed to be axisymmetric.  The radial bins
are the same as in the one-dimensional simulations: 120 cells covering 2.5 pc to 250
kpc where each cell is 10\% larger than the previous cell.  Requiring
the cells to have an aspect ratio of unity gives 30 angular
cells.  We exclude the region from 0 to 0.05 radians near either pole
because cell volumes go to zero due to the coordinate singularity there.

The simulations are not limited by the total volume of computation
required, but rather by the time to solution.  The
Courant--Friedrich--Levy (CFL) condition
for the central grid cells is quite severe: time steps are $\simeq$
10 years, and we would like to run the simulation for 12 Gyr.
Numerical schemes which allow the time step to increase with increasing
radius could be implemented, but the estimated speedup is quite
limited.  For a single processor this would save
a factor of $N_r(1-(r_i/r_o)^{1/N_r}) \simeq 10$ where $N_r$ is the
number of grid cells in the $r$-direction, $r_i$ is the innermost
radius, $r_o$ is the outermost radius, we have assumed log-spaced bins
in radius, and the numerical value is for our chosen parameters of
$r_i/r_o = 10^{-5}$ and $N_r=120$.  In a multi-processor environment,
the reduction in wallclock-time is given by the same formula with $r_i,
r_o$, and $N_r$ referring to the cells on a given processor rather than
the whole simulation.  For our simulations running on eight processors,
the speedup would be only $\simeq 40\%$.  

As the mass of the central SMBH increases, one might think that the
Bondi radius increases so that a moving inner grid could help in
speeding up the simulations. As discussed above, this will not work
since the gas losses from the evolving stars near the SMBH are
thermalized at the local value of the velocity dispersion, which
increases with BH mass.  In practice, the location of the ``Bondi
radius'' remains approximately fixed for the entire time of the
simulation, and varies chiefly due to fluctuations in the local speed
of sound caused by feedback from the SMBH and by the accretion of cold
gas.

Our requirements for the spatial resolution near the center of the
simulation, the amount of time for which the simulation must run, and
the CFL condition near the center have constrained us to use a rather
small number of cells.  We have performed a few simulations where the
number of grid cells is doubled in each dimension, and the simulation
ran for $\sim 1.3$ Gyr rather than 12 Gyr.  Most physical quantities
did not change significantly.  The exception was the change in the
BH mass, which was a factor of two larger.  Smaller cells allow
denser blobs and filaments to form, which have an easier time making
their way to the center of the simulation.  It is also important
to note that the size of the cells grows linearly with radius so that
physical structures of constant size (e.g., molecular clouds) will be
resolved progressively more poorly further from the center of the
simulation.  

\subsection{Physics}

For the most part the input physics is the same as that described in
\citet{ciotti:10} with a few exceptions described in detail below.
A complete description of the input physics is given in
\citet{ciotti:07}, \citet{ciotti:09-feedback}, and \citet{sazonov:05}.  Here we repeat the most
important aspects.  

The total gravitational potential of the model galaxy is assumed to be
a singular isothermal sphere plus a point mass for the central BH.  
This is good agreement with recent observations of the total
mass profile of early-type galaxies \citep{gavazzi:07, gavazzi:08}.
For simplicity we maintain this model in to the smallest radii,
although more complicated models may be more appropriate inside of a
fraction of the half-light radius.  The velocity dispersion parameter
of the isothermal potential is 260 km s$^{-1}$.  The gas is not self-gravitating.

The stellar distribution is given by a Jaffe profile with a total
mass of $3\times 10^{11} M_\odot$ and a projected half-mass radius of
6.9 kpc.  The mass-to-light ratio is assumed to be spatially constant
and is equal to 5.8 in solar units in the $B$ band at the present
time.  

Energy and mass input due to stellar evolution, type Ia supernovae,
star formation, type II supernovae are exactly the same.  The specific
energy of the material provided by stellar evolution is calculated
according to the solution of the Jeans equation for the given stellar
density distribution and gravitational potential as given in
\citet{ciotti:09-dynamics}.  Gas heating and cooling due to atomic
processes as well as Compton heating and bound--free absorption of
radiation from the AGN are calculated using the formulae from
\citet{sazonov:05}, as in \citet{ciotti:10}.

Gas temperatures are bounded from below by the fact that our atomic
cooling curve has an exponential cutoff below $10^{4}$ K and from above
by our assumed broad-line wind velocity of 10,000 km s$^{-1}$,
corresponding to $2.5\times 10^{9}$ K if all of the kinetic energy is
converted to thermal energy.  However, this very hot gas will always
be strongly outflowing until it distributes its energy to a much
larger mass of gas so that the temperature is of order the virial
temperature of the halo.  The upper temperature bound for infalling 
gas is given by the maximum temperature to which the AGN
photons can heat the gas.  This is given by the Compton temperature,
which we take to be 2.1 keV = $2.5\times 10^7$ K \citep{sazonov:05}.

Star formation is implemented using the standard Schmidt--Kennicut
prescription.  The star formation rate density is given by
\begin{equation}
  \dot{\rho}_* = \frac{\eta_{\rm form} \rho }{\tau_{\rm form}} \, ,
\end{equation}
where $\rho$ is the local gas density, $\eta_{\rm form}=0.1$ is a
dimensionless parameter for the rapidity of star formation and 
\begin{equation}
  \tau_{\rm form} = \max(\tau_{\rm cool}, \tau_{\rm dyn})
\end{equation}
\begin{equation}
  \tau_{\rm cool} = E/C
\end{equation}
\begin{equation}
  \tau_{\rm dyn} = \min(\tau_{\rm jeans}, \tau_{\rm rot})
\end{equation}
\begin{equation}
  \tau_{\rm jeans} = \sqrt{\frac{3\pi}{32 G \rho}}
\end{equation}
\begin{equation}
  \tau_{\rm rot} = \frac{2\pi r}{v_c(r)} \, ,
\end{equation}
where $E$ is the internal energy per unit volume of the gas, $C$ is
the volumetric cooling rate, $G$ is the Newtonian gravitational
constant, $r$ is the distance from the center of the galaxy, and
$v_c(r)$ is the circular velocity as a function of radius.  Type II
supernovae then return mass and energy to the ISM such that the mass
returned is 20\% of that processed through stars and the energy is
$E_{\rm SN} = 4\times 10^{-6} \dot{m}_{\rm SF} c^2$.  These values for the
energy and mass returned result from our assumption of a Salpeter
initial mass function with a low-mass cutoff of 0.1 $M_\odot$,
together with our assumption that every star above $8 M_\odot$ injects
its entire mass and $10^{51}$ erg into the ISM.  

In the momentum equation, the force exerted on gas due to Compton
scattering of photons from the AGN is treated as in the one-dimensional code.  Note
that because the vast majority of the photons from the AGN have
energies far below the electron rest mass energy, Compton scattering takes
place in the coherent limit where photon energies and numbers are
preserved.  Therefore, in both the optically thin and optically thick
limits, the force per unit mass on a fluid element is
\begin{equation}
  F/m = \frac{\kappa_{\rm ES} L(r)}{4 \pi r^2 c} \, , 
\end{equation}
where $\kappa_{\rm ES}$ is the electron scattering opacity and $L(r)$
is the total luminosity emitted interior to $r$ (that is, there
is no $e^{-\tau}$ factor, where $\tau$ is the optical depth).  

The force exerted by absorption of AGN photons by atomic lines is also
the same as in the one-dimensional code: $dp/dt = l/c$, where $l$ is the energy
absorbed per unit time in a given cell.  In the work reported on in
this paper, the radiation momentum associated with dust absorption
\citep[cf.][]{debuhr:11} is not included in the two-dimensional
calculation although it is included in the one-dimensional treatment.

Some physical processes are modeled using the same basic assumptions
as in the one-dimensional code, but two-dimensional simulations allow us to improve
the implementation.

This is the case with the BAL wind.  The one-dimensional and two-dimensional simulations use the
same formulae to compute the total mass, momentum, and energy
injected into the ISM for a given SMBH accretion rate.  However, the
one-dimensional simulations use a prescription based on pressure balance between
the outgoing wind and the ambient gas in the galaxy in order to
compute how the mass, energy, and momentum are radially distributed.
In the two-dimensional simulations, we can adopt the more realistic treatment of
simply injecting the desired mass, energy, and momentum into the
innermost radial cells and self-consistently computing the radial
transport of these quantities.

The one-dimensional code specifies the opening angle of the wind for the purpose of
solving for the radial distribution of the deposited
material.  The two-dimensional code requires a more precise specification of the
dependence of the outflowing material on angle from the pole.  We
choose the simple parameterization: 
\begin{equation}
  \frac{dq}{d\theta} = \frac{(n+1) \, \sin\theta \,
  |\cos^n\theta| \, Q}{4\pi}  \, ,
\end{equation}
where $q$ is a conserved quantity (mass, energy, or radial momentum),
$Q$ is the total amount of the conserved quantity to be injected.
We solve for $n$ such that the angle $\theta$ enclosing half of the
conserved quantity is the same as the opening angle specified in the
one-dimensional code.  For the fixed-opening-angle models considered here, $n=2$,
so that the half-opening angle enclosing half of the energy is $\sim$
$45^\circ$.  In terms of solid angle, this means that the wind is
visible from $\sim 1/4$ of the available viewing angles.  This
fraction is in agreement with observations of the fraction of obscured
and unobscured AGNs under the assumption that the two populations are
made up of a single population of objects that differ only in viewing
angle.

Finally, some processes of minor importance have been omitted from the two-dimensional
simulations so far.  This includes the prescription for star formation
and stellar remnant formation in the central sub-grid SMBH accretion
disk. The two-dimensional code implements no star formation in the disk---all
gas that flows in eventually either flows into the SMBH or back
onto the simulation grid as a conical wind.

The major difference between the calculations presented here and the
one-dimensional models is that the two-dimensional models do not yet implement any optically
thick radiative transfer algorithms.  We assume that the gas is always
optically thin at all wavelengths at all times.  There is also no
treatment of the temperature dependence of dust opacity.  Finally, the
radiation pressure due to stars formed in the simulation is also
omitted from the momentum equation.  

Experiments using one-dimensional models have shown that omitting the radiation
pressure on dust does not greatly affect the overall SMBH mass growth
when these other feedback mechanisms are included (less than a factor
of two).  However, star formation can be affected because infalling cold
shells can have their falling time extended by radiation pressure on
dust, allowing more stars to form as the gas is falling.  Improving
the treatment of radiative transfer in the two-dimensional code will be the subject
of future work.

Computational efficiency requires that we limit the velocities and
sound speeds in the simulation to prevent the CFL
condition from restricting the time step too dramatically.  At the high
end, we impose a limit of 30,000 km s$^{-1}$ on the gas velocities and sound
speeds.  The source term with the highest specific energy is the
mechanical wind due to SMBH accretion, which injects gas at 10,000 km s$^{-1}$.
The limit is three times higher, corresponding to 10 times the energy
density, so this limit should not greatly affect overall evolution of
the system.

In the one-dimensional simulations, we imposed the commonly used requirement that
the ISM temperature be above $10^4$ K.  This was
necessary because the one-dimensional code uses a fully explicit time integration
scheme and low thermal energies imply exceedingly short cooling times.
In the two-dimensional calculations, we use a semi-implicit time integration
method, making it possible to set the lower bound on temperature at a
lower value.  We impose three limits dictating that the gas does not drop
below the temperature of the cosmic microwave background, the
effective temperature associated with the AGN radiation field or the
effective temperature of the stellar radiation field.  The two-dimensional
simulations make use of the same cooling curve as the one-dimensional simulations,
which has an exponential cutoff below $10^4$ K.
Therefore, gas cannot reach these low temperatures by cooling alone,
but must undergo a dramatic expansion.  This occasionally occurs due
to the violent gas motions near the center of the simulation.  

We calculate the temperature of the cosmic microwave background
assuming an Einstein--de Sitter universe with an age of 13.7 Gyr where
the beginning of the simulation corresponds to 1.7 Gyr after the big
bang.  The difference between an Einstein--de Sitter universe and the
standard $\Lambda$CDM cosmology is minimal for our present purpose.

The effective temperature of the photons from the AGN is just
\begin{equation}
  T = \left(\frac{L}{4 \pi c a r^2}\right)^{1/4} \, , 
\end{equation}
where $a=8\pi^5k_B^4/15 c^3 h^3$ is the radiation constant.

It is easy to show that for a spherically symmetric distributed
luminosity source, the energy density in the photon field at $R$ is
\begin{equation}
  e(R) = \int_0^\infty \frac{j(r) \, dr}{2 c R} \log{\frac{R + r}{|R -
      r|}} \, ,
\label{eqn:en-dens}
\end{equation}
where $j(r)$ is the luminosity per unit volume.

However, we adopt the expression
\begin{equation}
  e(R) = \int_0^\infty \frac{r^2 j(r) \, dr}{c(r^2 + R^2)} 
\label{eqn:en-dens-approx}
\end{equation}
because it gives simple analytic integrals and has the correct form in
the limits when $r\gg R$ and $r \ll R$.  The integrand of Equation
(\ref{eqn:en-dens-approx}) underestimates the integrand of
Equation (\ref{eqn:en-dens}) when $r$ is near $R$.  However, the difference
between the two expressions is less than 30\% provided that $r$ and
$R$ differ by at least a factor of two.  Given that $l(r)$ is typically
a steeply falling function of $r$, the integral is not typically
dominated by the region where $r \sim R$.

For the present case of a Jaffe profile assuming a constant stellar mass-to-light ratio, this
gives
\begin{equation}
  e(R) = \frac{L}{4\pi c R^2} \frac{y (y^2 - 2 y \log y - 1)}{(y-1)^3}
  \, , 
\end{equation}
where $y=R/r_*$, $r_*$ is the scale radius from the Jaffe profile, $L$
is the total luminosity, and the expression has a finite limit when
$y=1$.  

There are also some minor computational differences between the two
codes.  The one-dimensional simulations use a fully explicit algorithm for time
integration.  In the two-dimensional code, the time integration is mostly explicit,
but we use a semi-implicit scheme for radiative cooling because the
cooling timescales can become very short.  If the time step required by
the CFL condition is shorter than the cooling time, the code updates
gas temperatures due to radiative cooling explicitly.  If the cooling
time is shorter than the CFL time step, the code evaluates the cooling
function at the advanced time, using a bisection algorithm to find the
root of the resulting equation for the gas temperature at the advanced
time.  

\subsection{AGN Feedback Parameterization}
The BAL originating near the SMBH provides energy, momentum, and
mass from a wind to the ISM according to the equations:

\begin{equation}
  \dot{M}_{\rm BH} = \dot{M}_{\rm infall}/(1+\eta) \, , 
  \label{eq:mdot-bh}
\end{equation}
\begin{equation}
  \eta = 2 \epsilon_W c^2 / v_W^2 \, , 
  \label{eq:eta}
\end{equation}
\begin{equation}
  \dot{E}_W = \epsilon_W \dot{M}_{\rm BH} c^2 \, , 
  \label{eq:edot}
\end{equation}
\begin{equation}
  \dot{P}_W = 2 \epsilon_W c^2  \dot{M}_{\rm BH} / v_W \, , 
  \label{eq:pdot}
\end{equation}
and
\begin{equation}
  \dot{M}_W = 2 \epsilon_W c^2  \dot{M}_{\rm BH} / v_W^2 \, , 
  \label{eq:mdot-w}
\end{equation}
where $v_W$ is the velocity of the BAL, taken here to be 10,000 km s$^{-1}$.  
As discussed in detail in \citet{ostriker:10}, these expressions
guarantee that the mass, energy, and momentum carried by the wind are
self-consistent.

\citet{ciotti:09-feedback} considered two classes of mechanical feedback,
denoted the A and B models.  For the A models, both the
mechanical efficiencies and the wind opening angles were independent
of the accretion rate.  The B models have mechanical efficiencies and wind
opening angles that varied with SMBH accretion rate such that both 
quantities are small at small accretion rate and reach a specified
maximum at the Eddington rate.

Independently of the mechanical feedback model, the radiative
luminosity of the AGN is given by
\begin{equation}
  L = \epsilon_{\rm EM} \dot{M}_{\rm BH} c^2  \, ,
  \label{eq:lum}
\end{equation}
where the electromagnetic efficiency is given by the advection
dominated accretion flow inspired
\citep{narayan:94} formula:
\begin{equation}
  \epsilon_{\rm EM} = \frac{\epsilon_0 A \dot{m}}{ 1 + A \dot{m}}
  \, ,
  \label{eq:epsilon-em}
\end{equation}
and $A=100$ and $\epsilon_0=0.1$.  The dimensionless mass accretion rate is
\begin{equation}
  \dot{m} = \frac{\dot{M}_{\rm BH}}{\dot{M}_{\rm Edd}} = 
  \frac{\epsilon_0 \dot{M}_{\rm BH} c^2}{L_{\rm Edd}} \, ,
\end{equation}
where $L_{\rm Edd}$ is the Eddington luminosity.

The mechanical wind efficiency is given by $\epsilon_W=$ constant
for A models.
The specific parameters for the simulations presented here are given
in Table \ref{tab:mydata}.  The present work considers two-dimensional
analogs of the simpler A models only.  

We also ran two-dimensional analogues of the \citet{ciotti:09-feedback} B models where the
mechanical feedback efficiency falls off at low Eddington ratios (as
expected from computational studies of the creation of BLWs in the
inner few hundred gravitational radii \citep{kurosawa:09-2d,
  kurosawa:09-efficiency, kurosawa:09-3d}).  In one dimension, mechanical feedback in
the B models was sufficient to regulate the SMBH growth.  However,
in two dimensions equivalent models were found to undergo too much SMBH growth and
had distributions of Eddington ratios very different from those
observed for actual galaxies.  Comparing one-dimensional and two-dimensional A models reveals
that feedback is in less effective in two dimensions: more energy is required for
the SMBH to effectively regulate its growth.  For two dimensions, the low
efficiencies at low Eddington ratio combined with the diminished
effectiveness of feedback in two dimensions make the two-dimensional B models 
observationally unacceptable.  

The model most similar to that used by \citet{dimatteo:05} is A0.
The model has an efficiency of $5\times 10^{-3}$, but we take into
account the fact that the wind carries not only energy but also
momentum and mass into the ISM.  Matching our momentum input rate to
that used by \citet{debuhr:11} gives $\epsilon_W = v_w
\epsilon_0 \tau / 2 c \simeq 0.04$, where $\tau = 25$ is their chosen
wavelength-averaged optical depth parameter and $v_w = 10,000$ km s$^{-1}$ is
our chosen BAL wind velocity parameter.  The present models do not
include a simulation with such a high feedback efficiency.  Our model
that most closely approximates those of \citet{debuhr:11} is
again A0, with the caveat that we inject less momentum for the same
BH luminosity.  Note that \citet{debuhr:11} effectively
assume a much lower wind velocity (not intended to model BAL winds)
and as a result the energy actually injected into the ISM in their
simulations is similar to that used in \citet{dimatteo:05}.  The above
considerations are intended to determine what parameters we
would need to adopt for our present simulations in order to produce
models similar to existing ones by other authors given our assumed BAL
velocity.  A closer study of the similarities and differences between
these AGN feedback models is the subject of a forthcoming paper.

\subsection{Angular Momentum}
The one-dimensional simulations did not permit the simulated gas to
have non-zero angular momentum.  The two-dimensional simulations assume axisymmetry
and compute the velocity in the $\phi$-direction.  We must assume an
angular momentum profile for the injected gas.  In the present
simulations, we avoid forming a rotationally supported gas disk by
choosing the radius of centrifugal support to be inside the innermost
grid cell.  

This assumption gives rotation velocities comparable to the slowest of
the SAURON slow-rotators \citep{emsellem:04, emsellem:07}.  Thus, our
model galaxy is representative of a rather small set of observed
galaxies.  However, the advantage of assuming very little rotation is
that it allows us to avoid specifying an ad hoc prescription for
angular momentum transport.  The low angular momentum case also allows
the closest comparison to the existing one-dimensional simulations, allowing us to
isolate the effect of dimensionality.  In future work we plan to
implement standard recipes for angular momentum transport, including
an $\alpha$ disk \citep{shakura:73} and models based on gravitational
torques \citep{hopkins:10-analytic-preprint}.  Taking angular momentum
transport into account is likely to have a significant effect on our
computed results.  Our consideration of the low angular momentum case
will facilitate isolating the effect of angular momentum in future analysis.

The assumed net specific angular momentum of the stars providing gas
in the simulation is thus taken to be given by
\begin{equation}
  \frac{1}{v_\phi(R)} = \frac{d}{\sigma R} + \frac{1}{f \sigma} + \frac{R}{j} 
   \, , 
\end{equation}
where $v_\phi$ is the velocity in the azimuthal direction, $R$ is the
distance to the $z$-axis, $\sigma$ is the central one-dimensional line-of-sight
velocity dispersion for the galaxy model, and $d$, $f$, and $j$ are
adjustable parameters controlling the angular momentum profile at
small, intermediate, and large radii, respectively.  This parameterization
gives solid body rotation at radii less than $d$ and constant specific angular
momentum of $j$ at large radii.  The second term on the right
prevents the rotational velocity from exceeding $f \sigma$ at any
radius.  If there is a range of values of $R$ for which the $f \sigma$
term dominates, then is a region of intermediate radii with constant rotational
velocity.  

\subsection{Initial Conditions}
The initial conditions are chosen to match those of the one-dimensional
simulations, fully described in \citet{ciotti:10}.  Briefly, the total
gravitational potential is that of a singular isothermal sphere with
one-dimensional velocity dispersion 260 km s$^{-1}$ and  with a central point mass
with initial mass $3\times 10^8 M_\odot$.  The stellar distribution is
a Jaffe profile with projected half-light radius 6.9 kpc,
mass-to-light ratio 5.8 in solar units, and total mass $3\times 10^{11}
M_\odot$.  These parameters are chosen to put the initial galaxy on
the Fundamental Plane \citep{djorgovski:87, dressler:87}, the
Faber--Jackson \citeyearpar{faber:76} relation, and the \citet{magorrian:98} relation.
All of the relevant dynamical properties of the galaxy models are
given in \citet{ciotti:09-dynamics}.  
The gas density is initially set to a very low value
so that the gas in the simulation comes almost exclusively from
explicit source terms arising from stellar evolution.

\subsection{Boundary Conditions}
We assume reflecting boundary conditions on the $\theta$ boundaries
occurring at either pole.  On the outer radial boundary we assume that
all fluid quantities are constant.  This allows both outflow and
inflow depending on the state of gas just inside the outer boundary.

On the inner radial boundary we assume reflecting boundary conditions if
the innermost radial velocity is positive.  Physical processes that
inject mass (such as the BAL wind) are handled as explicit
source terms acting in the first set of radial cells.  This allows us
to easily handle the cases of very strong, very weak, and intermediate
flows of energy, mass, and momentum onto the computational grid with a
single source term.  If the innermost radial velocity is negative, we
use an outflow (off of the computational grid toward the center of the
simulation) boundary condition where all fluid variables are constant
across the boundary.  The one exception in this case is the radial
velocity itself.  If the cell is inside the locally estimated Bondi
radius (that is, the Bondi radius has been resolved), then no limit is
imposed upon the inflow velocity.  However, if the cell is outside the
locally estimated Bondi radius (the Bondi radius is unresolved), then
the radial velocity is limited to the velocity that would result in
mass transport consistent with the Bondi accretion rate.  That is
\begin{equation}
  v_{r,0} = 
  \left\{
  \begin{array}{r@{\quad:\quad}l@{\quad}l}
    0 & \mbox{if} & v_{r,1} > 0 \\ 
    v_{r,1} & \mbox{if} & v_{r,1} < 0, r_1 < r_B \\ 
    (r_B/r_1)^2 v_{r,1} & \mbox{if} & v_{r,1} < 0, r_1 > r_B \, .\\ 
  \end{array}
  \right. 
  \label{eq:bondi-boundary-condition}
\end{equation}
As discussed above, we are careful to resolve the Bondi radius even
for gas heated to the maximum expected temperature for infalling gas,
the Compton temperature.  However, the BAL wind has an even higher
specific energy.  Typically, the BAL wind is flowing strongly outward
and the Bondi radius is always resolved.  However, it is possible for
the wind to fill the inner region of the galaxy with gas heated above
the Compton temperature.  In this case,  Equation
(\ref{eq:bondi-boundary-condition}) ensures that the SMBH accretion rate
is reduced accordingly.  

It is unfortunately not possible to cleanly separate the SMBH
scales from the galactic ones.  Physically, this is because the
source of gas in the galaxy is the stars, evolving on galactic
timescales.  Meanwhile, the SMBH is easily able to affect gas on
kiloparsec scales.  This ties the scales together in a feedback loop
that cannot be easily separated into two separate simulations, or a
simulation plus a sub-grid model.  Furthermore, the classic Bondi
solution is for the simple case of a point mass.  If the mass enclosed
by the Bondi radius is dominated by the galaxy rather than the central
BH, then the classic Bondi solution is modified, particularly
if the galactic mass is not spherically symmetric.

\section{Results}
\label{sec:results}
The character of the SMBH accretion changes dramatically in going from
one to two dimensions.  In one dimension, accretion events happen when a cold
shell of gas forms at $\simeq$ 100 pc.  The shell falls into the SMBH as
a unit and, after a series of sub-bursts in which direct and
reflected shock waves interact and carry new material for accretion on
the SMBH, triggers a dramatic release of energy from the SMBH.  This
leaves a sphere of hot gas at the center of the simulated galaxy.
Subsequent cold shells can only reach the center when the gas beneath
them either cools or is compressed inside the innermost grid point.  The
hot gas generated by radiative and mechanical AGN feedback is able to
prevent SMBH accretion until it cools.  These processes lead to dramatic
bursts followed by long, extremely quiet periods, spaced by the cooling
time of the central gas.  As the collapsing cold shell sits on top of
hot and low density gas, it was already clear from the previous one-dimensional
simulations that the shells are RT unstable.  

In two dimensions, cold gas forms again at $\simeq$ 100 pc.  However, cold gas
takes the form of rings rather than shells due to the classical RT
instability.  Shells sometimes form, but they fragment quickly.  If
there is hot gas beneath the cold ring, both the RT and
convective instabilities operate to allow the cold gas to fall to the
center of the simulation unimpeded.  These instabilities cannot
operate in one dimension.  In two dimensions, they allow both higher and more
chaotic SMBH accretion rates.

\begin{figure*}
\centering
\includegraphics[width=0.85\textwidth]{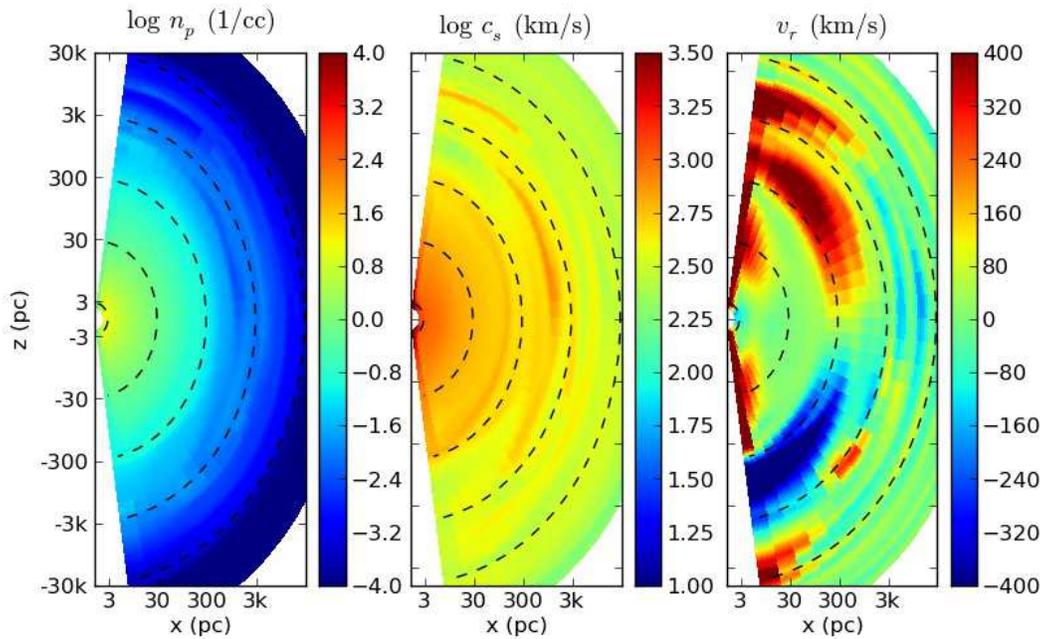}
\caption{Simulation snapshot during a quiescent period.  On the left, gas
    density in number of protons per cubic centimeter.  In the center,
    log sound speed in kilometers per second.  On the right, the
    radial velocity in kilometers per second.  The $x$- and $y$-axes
    are logarithmic in the distance to the SMBH.  Outflowing
  gas from past accretion events is visible from 300 pc to 10 kpc.
  Persistent low-level accretion maintains a narrow continuous outflow near the poles.}
\label{fig:snap-quiescent}
\end{figure*}

\begin{figure*}
\centering
\includegraphics[width=0.85\textwidth]{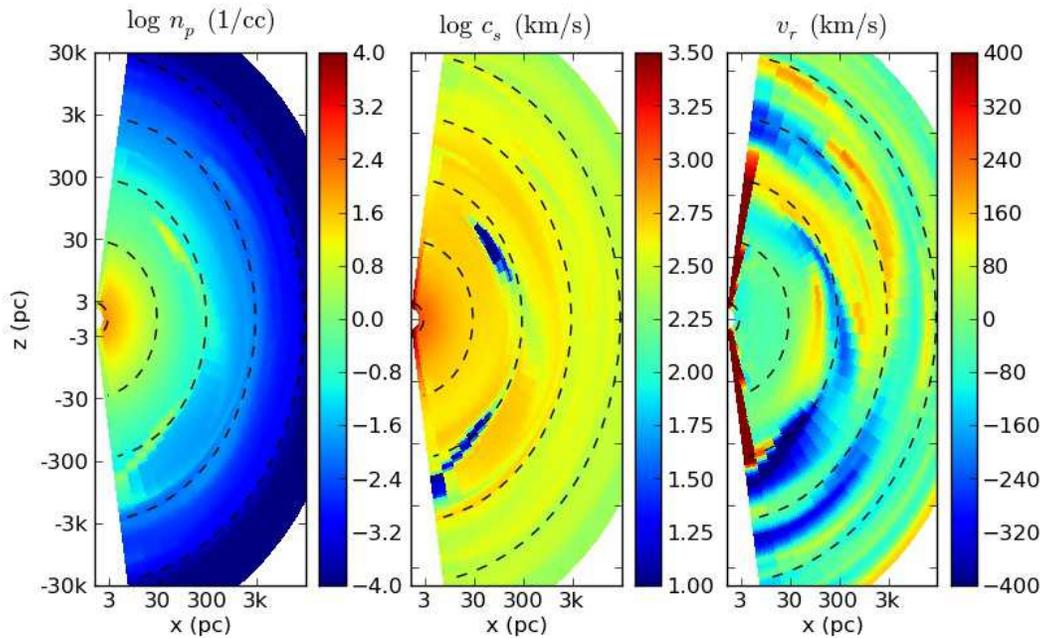}
\caption{Snapshot from an axisymmetric simulation showing a cold
    blob falling to the center of the galaxy.  The cold gas
    was produced by enhanced cooling in an overdense quasi-spherical
    shell with a covering fraction of about one-third of the sphere.
    The gas quickly collapses to a ring with a small covering fraction
    and/or fragments as it freely falls to the center of the
    simulation.}
\label{fig:snap-before}
\end{figure*}

\begin{figure*}
\centering
\includegraphics[width=0.85\textwidth]{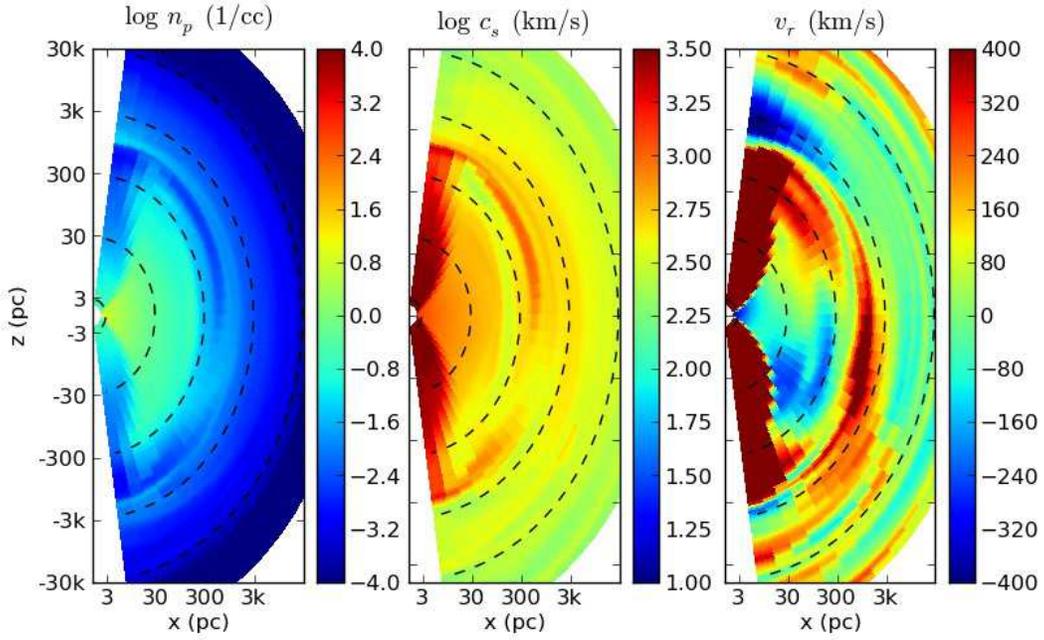}
\caption{Simulation snapshot during an accretion event.  A significant
  quantity of hot, outflowing gas injects energy and momentum into the
  interstellar medium at $r \simeq 1$ kpc.}
\label{fig:snap-during}
\end{figure*}

\begin{figure*}
\centering
\includegraphics[width=0.85\textwidth]{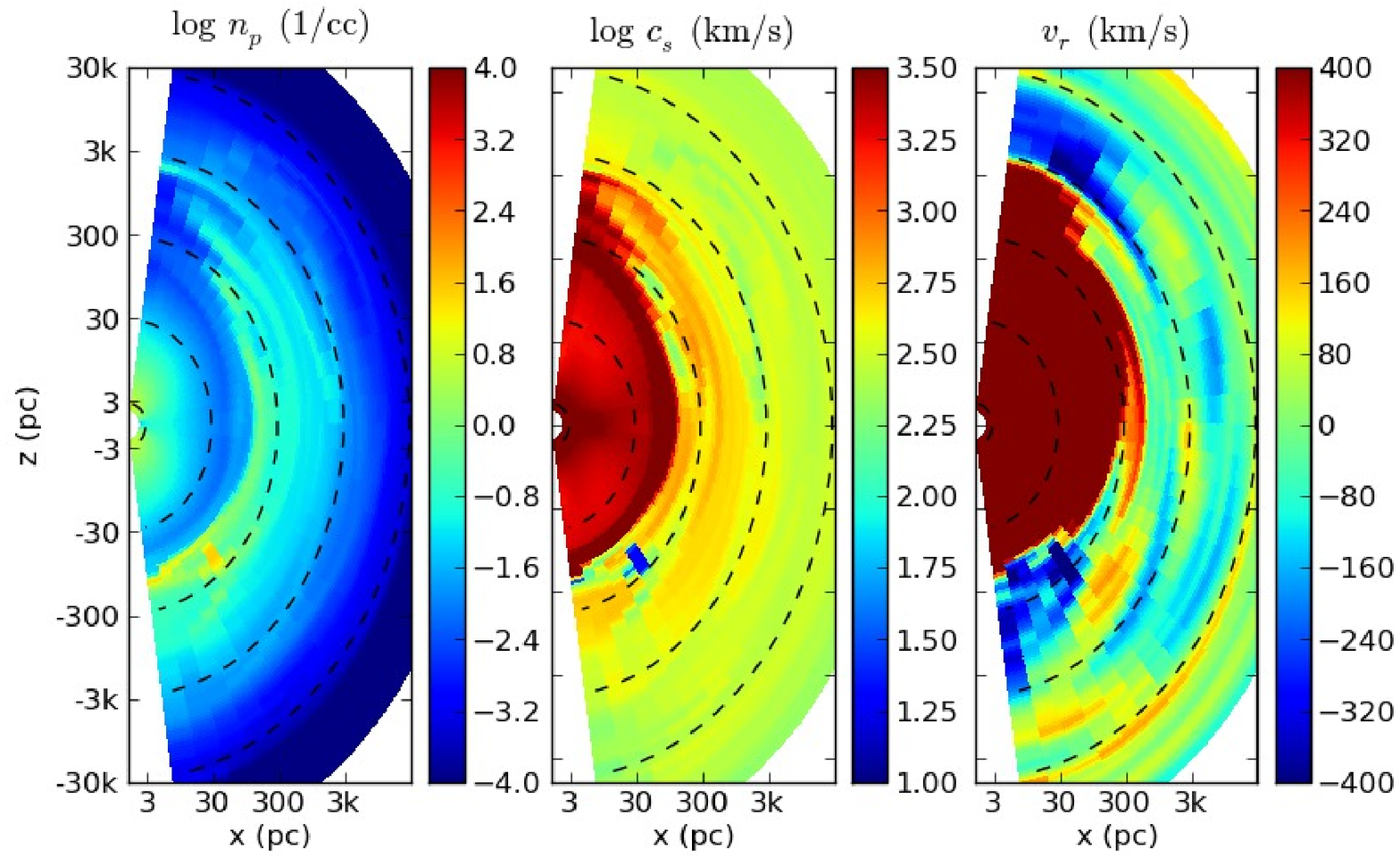}
\caption{Simulation snapshot showing the final stages of a major
  accretion event.  A hot, expanding bubble of gas extends to 100 pc,
  shutting down further SMBH accretion.  Dense, overlying gas has caused
  the initially bipolar BAL wind to become quasi-spherical.  The hot
  bubble is breaking through the overlying gas in at the north pole,
  which will lead to a unipolar wind.}
\label{fig:snap-after}
\end{figure*}

Figure \ref{fig:snap-quiescent} shows the simulation during a
relatively quiet period.  Figure \ref{fig:snap-before} shows the start
of an accretion event.  A cold blob of gas is freely falling to the
center of a two-dimensional simulation.  
Figure \ref{fig:snap-during} shows a
simulation snapshot just after an accretion event with the bipolar BAL
wind flowing away from the center of the simulation.  Gas is able to
continue to fall into the SMBH via the simulation midplane.  Finally, Figure
\ref{fig:snap-after} shows a simulation snapshot significantly after
an accretion event (but not so long that the galaxy is able to return
to a quiescent state).  Dense overlying gas has caused the BAL wind to
become nearly isotropic, making additional accretion via the midplane
impossible.  

\subsection{Black Hole Growth}

\begin{figure*}
\centering
\includegraphics[width=0.7\textwidth]{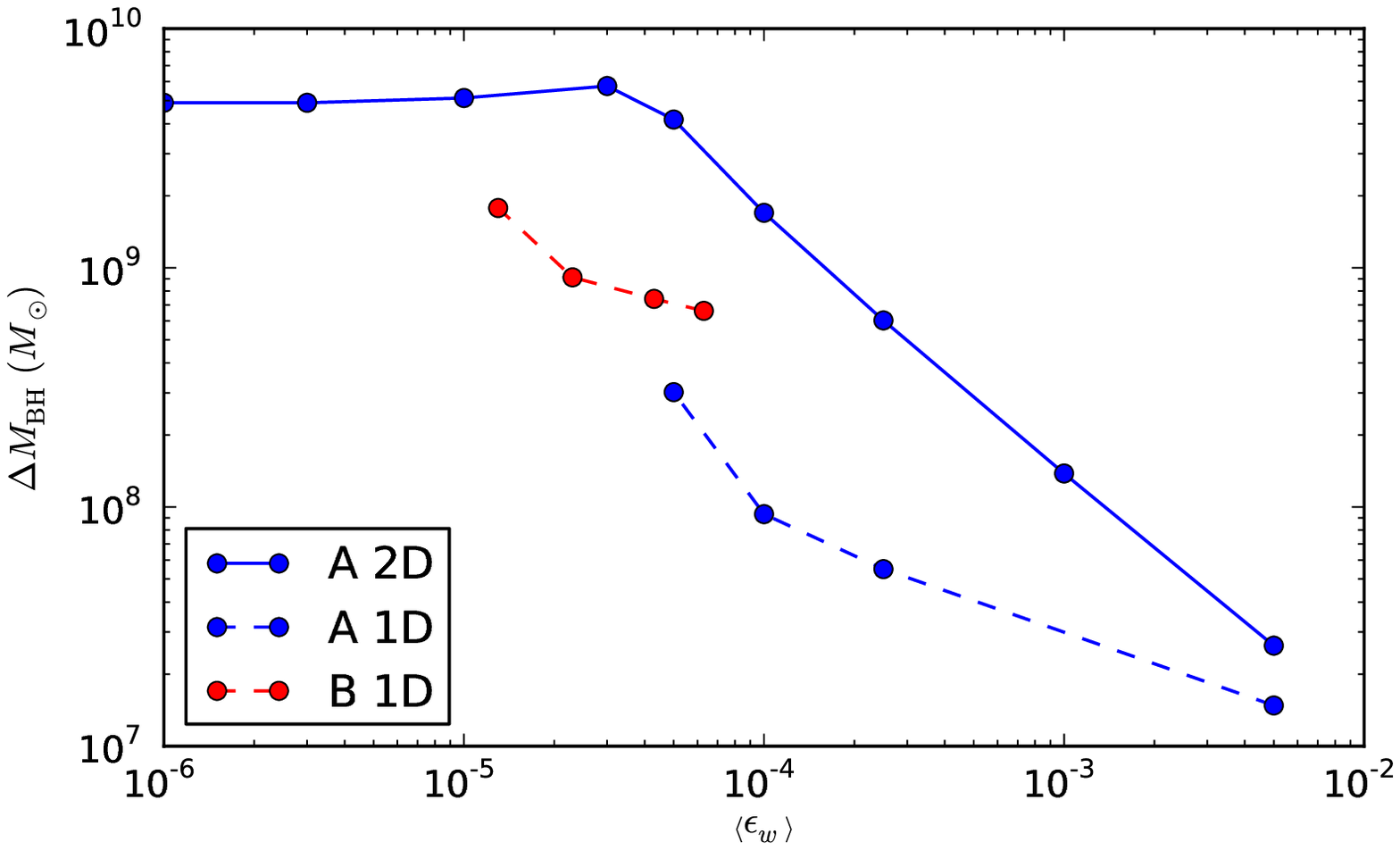}
\caption{Final SMBH masses vs. mass-averaged wind Efficiency for one-
  and two-dimensional A models.  The low-efficiency two-dimensional models
  produce SMBHs at the upper end of the range of observed central SMBHs
  for the characteristics of our model galaxy.  Two-dimensional models
  allow more SMBH growth at a given efficiency because instabilities
  allow cold shells of gas to break up and fall into the center of the
  simulation with greater ease than in the one-dimensional case.  A
  more accurate two-dimensional treatment of the radiative transfer in optically
  thick regions would probably lead to a reduction in the SMBH masses.}
\label{fig:bh-vs-eff}
\end{figure*}

Figure \ref{fig:bh-vs-eff} shows SMBH mass versus mechanical feedback
efficiency for one- and two-dimensional A models as well as
one-dimensional B models.  The SMBHs undergo more growth in two
dimensions at a fixed feedback efficiency owing to RT instabilities,
making it easier for cold gas to fall in.

\subsection{Time Dependence of SMBH Accretion}

The character of the time dependence of SMBH accretion is much different
in two dimensions than in the one-dimensional case.  In one dimension, SMBH accretion occurs in a few
bursts well separated in time.  Occasionally a given burst will have a
complex character, being composed of many sub-bursts.  In two
dimensions, there are still occasionally events that can be
characterized as bursts followed by quiescent periods.  However, the
quiescent periods are shorter and the SMBH is more active during them
compared to the one-dimensional case.  Furthermore, during times far from a major
burst when gas is building up in the galaxy, the SMBH is far more active
than in the one-dimensional case and the accretion takes on a stochastic character.

\begin{figure*}
\centering
\includegraphics[width=0.7\textwidth]{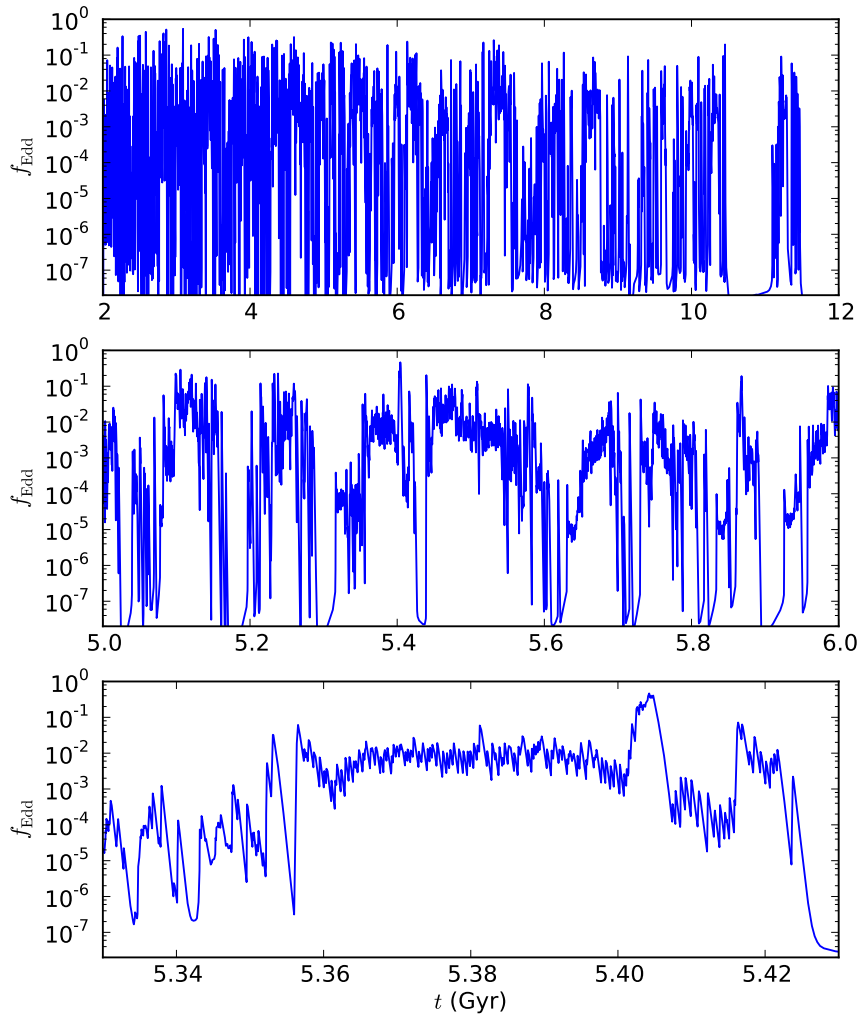}
\caption{Eddington ratio as a function of time, for three different
  time intervals in the $A2$ simulation.}
\label{fig:eddington-ratio}
\end{figure*}

Figure \ref{fig:eddington-ratio} shows the Eddington ratio $L/L_{\rm
  Edd}$ as a function of time for part of the $A2$ simulation.
The SMBH accretion is much more chaotic than in the one-dimensional case.  In the
one-dimensional case, a cold shell forms and falls in as a unit only
after the gas interior to it has also cooled.  In two dimensions,
these same cold shells form, but they immediately break up into blobs
that are RT unstable.  The cold blobs fall in to the SMBH
on a free-fall time in a much more disorganized fashion compared to
one dimension.

\begin{figure*}
  \centering
  \includegraphics[width=0.7\textwidth]{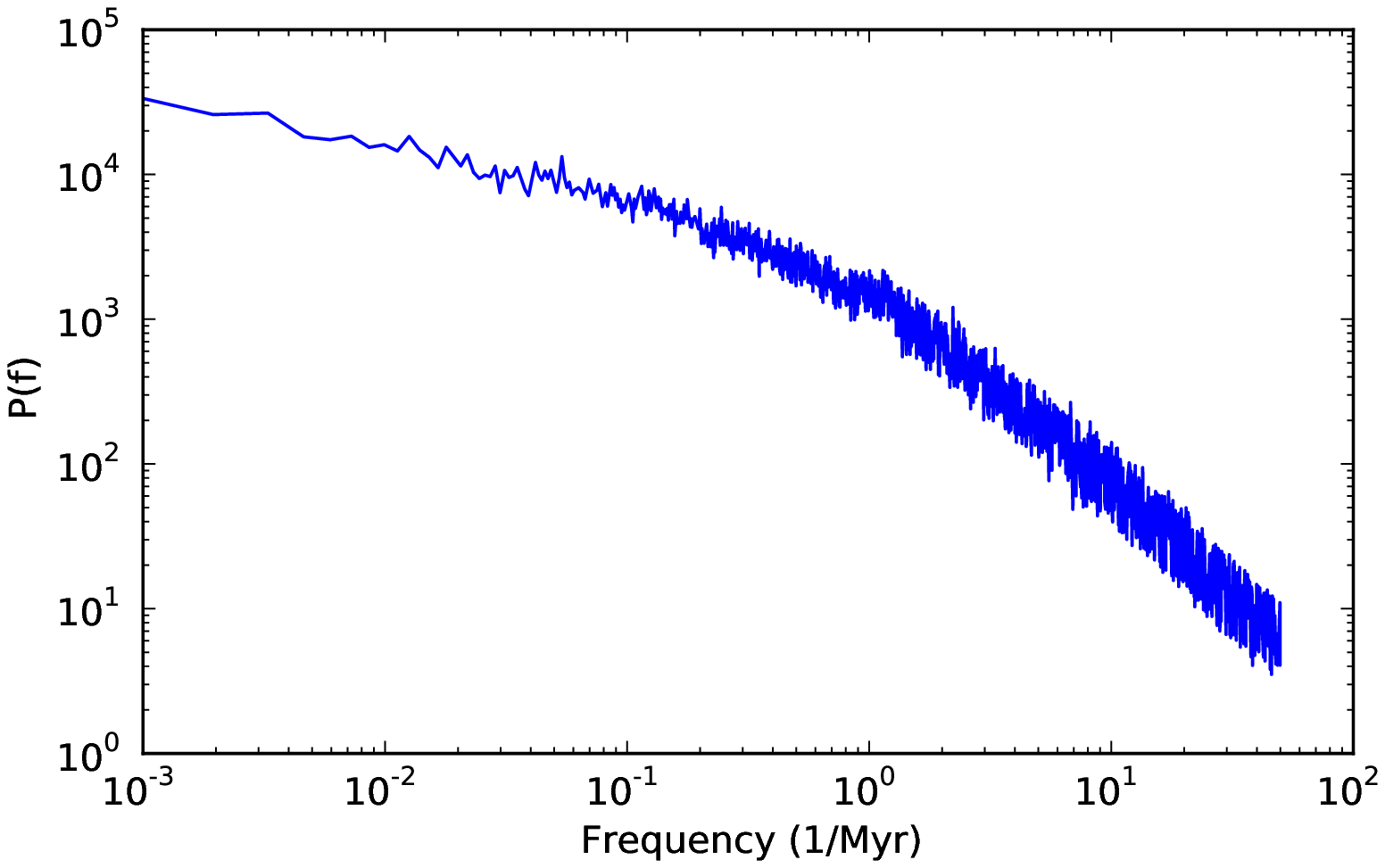}
  \caption{Power spectrum of the dimensionless mass accretion rate
    $\dot{M}_{\rm BH}/\dot{M}_{\rm Edd}$ for A2.  The units of the 
    $y$-axis are arbitrary.  For frequencies higher than the inverse of
    several times the central accretion disk timescale, the power
    spectrum is proportional to $1/f$, the frequency response of the
    low-pass filter applied by our central accretion disk.  For lower
    frequencies, the power spectra are determined by the physics of
    gas input, cooling, and feedback.  At these low frequencies, the
    power spectrum is nearly flat: $\propto f^{-1/4}$.  That is, the
    accretion onto the SMBH has a power spectrum nearly the same as
    white noise.}
  \label{fig:power-spectrum}
\end{figure*}

Figure \ref{fig:power-spectrum} shows the power spectrum of the
Eddington ratio as a function of time for the $A2$ simulation.  In
spite of the complex, detailed implementation of the physics of
stellar evolution, atomic cooling, star formation, supernovae, and AGN
feedback, the power spectrum reveals a simple, gently sloped power law
(roughly frequency to the one-fourth power) at low frequencies.  At
high frequencies, the power spectrum falls off as $1/f$ because of the
filtering effect of the SMBH accretion disk.  

\begin{figure*}
\centering
\includegraphics[width=0.7\textwidth]{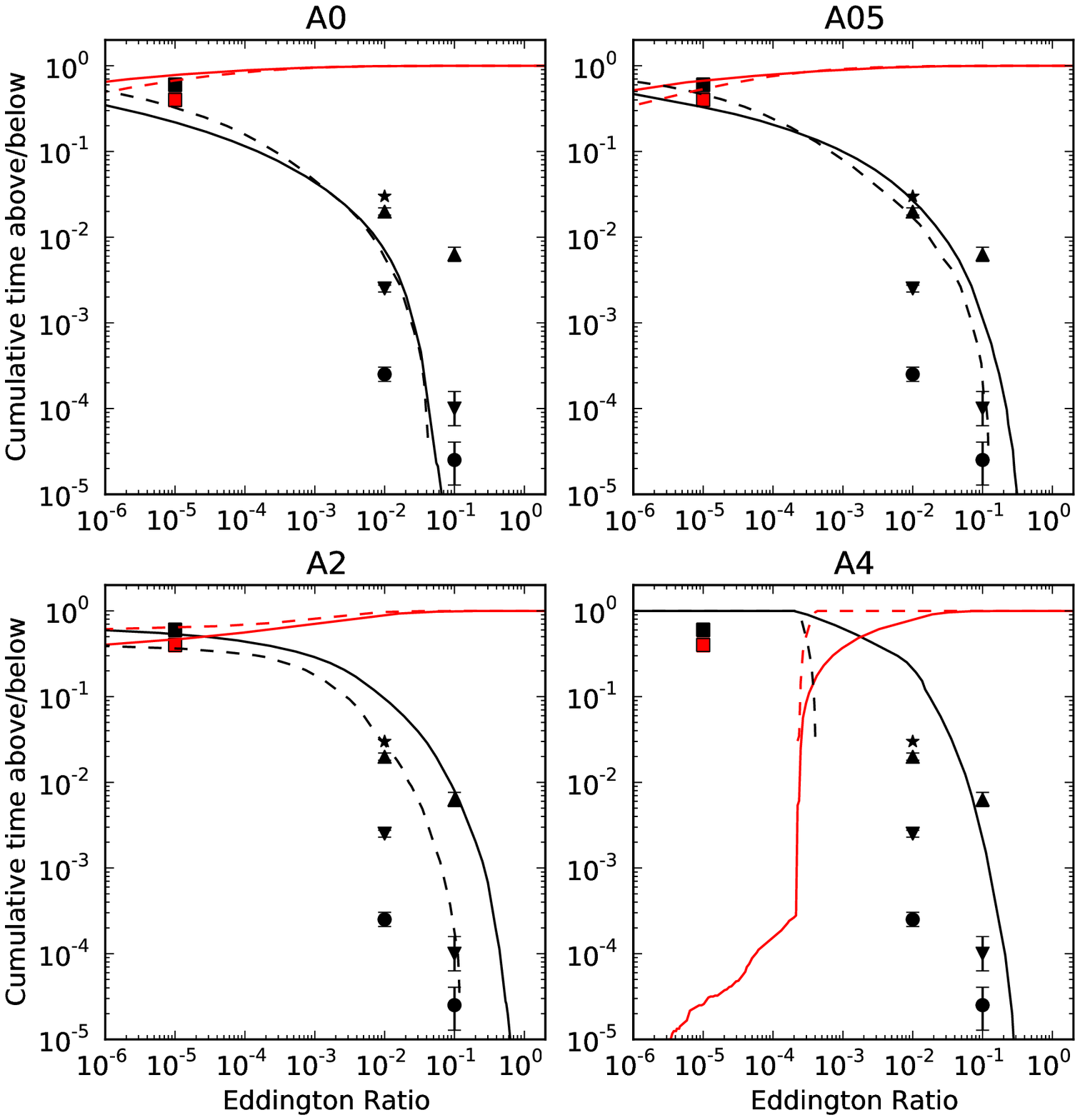}
\caption{AGN duty cycle as a function of Eddington ratio for the A2
  simulation.  The black/red lines show the cumulative time
  above/below the given Eddington ratio.  Solid lines are computed
  using the entire simulation time, while dashed lines use only the
  final 2 Gyr.  Points are observational constraints and are similarly
  colored black or red according to whether they are measurements of
  the fraction of objects above or below the given Eddinton ratio.
  Downward-pointing triangles are from \citet{heckman:04}, circles
  from \citet{greene:07}, upward-pointing triangles from
  \citet{kauffmann:09}, and squares are from \citet{ho:09}.  All of
  these use low-redshift observations and should thus be compared to
  the dashed lines.  The star is a constraint from \citet{steidel:03}
  for the Lyman-break galaxies at high redshift showing nuclear
  activity, and therefore should be compared to the solid lines.
  There is significant observational disagreement about the fraction
  of $3\times10^8 \, M_\odot$ BHs accreting at 1\% or 10\% of
  the Eddington rate, even at low redshift.  Nevertheless, the A05 and
  A2 simulations fit both the high-redshift and low-redshift
  constraints reasonably well given the state of the observational
  data.}
 \label{fig:agn-duty-cycle}
\end{figure*}

Figure \ref{fig:agn-duty-cycle} shows the cumulative distribution of
Eddington ratios for the A2 simulation.  The black (red) lines show
the cumulative time the simulation spends above (below) the given
Eddington ratio.  Solid lines show the cumulative distribution of
Eddington ratios for the entire simulation, while the dashed lines
show the distribution for the final 2 Gyr (corresponding to low
redshifts).  Low-redshift observational constraints for the fraction
of $3 \times 10^{8} \, M_\odot$ BHs accreting at 1\% and 10\%
of the Eddington rate are taken from \citet{heckman:04}, \citet{greene:07},
  \citet{ho:09}, and \citet{kauffmann:09}.  A constraint on the fraction of high-redshift
Lyman-break galaxies showing nuclear activity is taken from
\citet{steidel:03}.  

There is reasonable agreement between the simulations and
observational constraints for a mechanical efficiencies $\epsilon_W$
between $10^{-3}$ and $10^{-4}$.  This range is in agreement with the
recent study by \citet{arav:11-preprint}.  They presented new observations as
well as values drawn from the literature for $\dot{E}$ and $\dot{M}$
for five broad-line systems.  The mean value of $\log \epsilon_W$ for
the five systems is $-4.2$ with a mean error of 0.2 dex for each system
and a standard deviation of 0.4 dex for the five points.

\subsection{Disk Wind Opening Angle}
The opening angle of the disk wind is expected to have an effect on
the SMBH growth by controlling the effectiveness of the AGN feedback.
At very small opening angles, the energy injected by the SMBH drills
through the nearby gas and exits the central region as a narrow beam.
All of the energy is eventually deposited at radii beyond one kpc,
leaving the cooling and infalling gas in the central region
essentially undisturbed.  This should result in large SMBH growth.

As the opening angle increases, one might expect that mechanical
feedback would become more and more effective as the wind interacts
with an increasingly large fraction of the gas surrounding the SMBH.
This would imply that isotropic feedback would result in the least
SMBH growth.

\begin{figure*}
\centering
\includegraphics[width=0.7\textwidth]{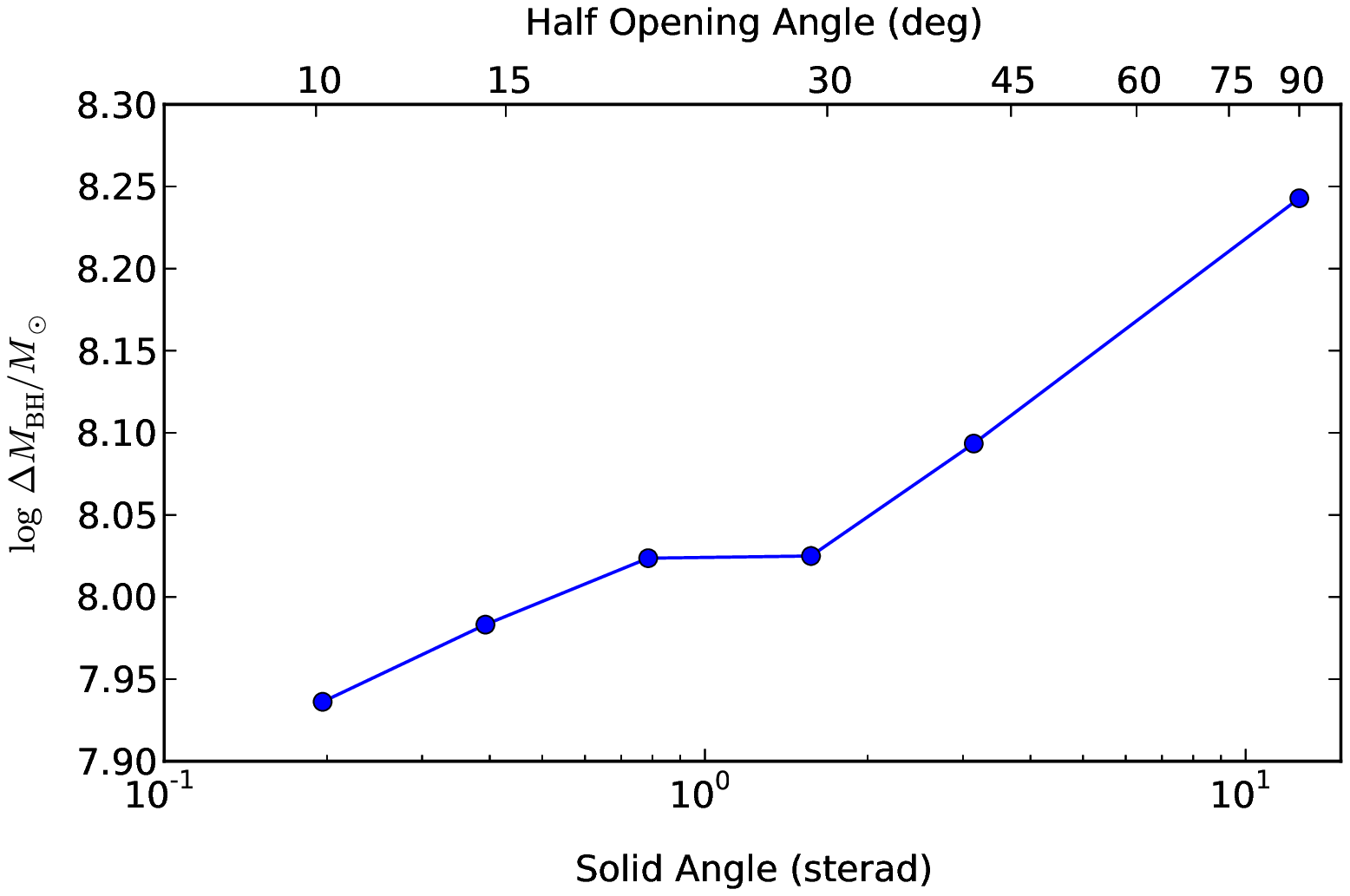}
\caption{Change in SMBH mass vs. wind opening angle for
  $\epsilon_W=10^{-3}$.  Mechanical feedback becomes less effective
  for opening angles larger than $45^\circ$, corresponding to a
  covering fraction of 1/4 of the sky.  Overall, the change in BH 
  growth is less than a factor of two for a wide range of wind
  opening angles. }
\label{fig:mbh-vs-angle}
\end{figure*}

Figure \ref{fig:mbh-vs-angle} shows final SMBH masses versus wind
opening angle.  The largest SMBH growth occurs when the wind is
isotropic.  A very wide opening angle allows the wind to effectively
couple to all of the gas at small radius, temporarily suppressing SMBH
accretion.  However, the wind's outward progress is soon stopped
because the wind must essentially lift the entire overlying gaseous
atmosphere.  The wind is not able to move a significant amount of gas
to a sufficiently large radius to prevent eventual accretion onto the
SMBH.

This implies that there is an opening angle that results in minimal
SMBH growth.  The angle will likely be a function of the mechanical
feedback efficiency as well as the other physical parameters for the
gravitational potential, stellar distribution, mass-loss rates, and so
forth.  Figure \ref{fig:mbh-vs-angle} shows the increase in SMBH
growth at large opening angles, but does not show evidence for the
expected increase in SMBH growth as the opening angle becomes very
small.  However, it is worth noting that the change in BH mass
growth is less than a factor of two for a wide range of wind opening angles.

\subsection{Star Formation and Galactic Winds, and Gas Content}

Table \ref{tab:mydata} gives total mass of stars formed, total mass of
gas driven beyond 10$R_e$, and final mass of gas within 10$R_e$ for
each simulation.  Star formation consumes about 30\%
 of the total mass budget in the two-dimensional simulations and is very
insensitive to the details of the AGN feedback.  This is in good
agreement with the one-dimensional simulations at low mechanical efficiencies.
However, at high feedback efficiencies, the one-dimensional simulations drive
significant quantities of gas out of the galaxy, leading to low star
formation rates and low final gas content.  In this respect the one-dimensional and
two-dimensional simulations disagree.  However, this is to be expected since
assuming spherical symmetry gives the most favorable situation for
turning a central energy source into a global outflow.  In two
dimensions, energy can escape via low-density channels and fail to
participate in driving an outflow.

\begin{deluxetable*}{lrrrrrrr}
\tablenum{1}
\tablecolumns{7} 
\tablewidth{0pc} 
\tablecaption{Properties of Computed Models}
\tablehead{
  \colhead{ Model} & 
  \colhead{ $\epsilon_W$} & 
  \colhead{ $\left< \epsilon_{\rm EM} \right>$} & 
  \colhead{ $\log\Delta M_{\rm BH}$} &
  \colhead{ $\log\Delta M_*$} & 
  \colhead{ $\log\Delta M_W$} &
  \colhead{ $\log M_{\rm gas}$}}
\startdata 
A0 & $5   \times 10^{-3}$ & 0.041 & 7.40 &  9.98 & 10.28 & 9.67 \\
A05& $           10^{-3}$ & 0.060 & 8.10 &  9.94 & 10.33 & 9.50 \\ 
A1 & $2.5 \times 10^{-4}$ & 0.067 & 8.71 &  9.88 & 10.35 & 9.45 \\ 
A2 & $           10^{-4}$ & 0.067 & 9.18 &  9.91 & 10.29 & 9.49 \\
A3 & $5   \times 10^{-5}$ & 0.063 & 9.55 &  9.94 & 10.17 & 9.58 \\
A4 & $3   \times 10^{-5}$ & 0.060 & 9.72 &  9.93 & 10.11 & 9.50 \\
A5 & $1   \times 10^{-5}$ & 0.048 & 9.71 &  9.98 & 10.03 & 9.50 \\
A6 & $3   \times 10^{-6}$ & 0.047 & 9.69 & 10.00 & 10.03 & 9.52 \\
A7 & $1   \times 10^{-6}$ & 0.047 & 9.69 &  9.99 & 10.03 & 9.52 \\

\enddata 
\label{tab:mydata} 
\tablecomments{Final properties of the simulated galaxies, where
  $\epsilon_W$ is the mechanical wind efficiency, $\left<
    \epsilon_{\rm EM} \right>$ is the mass-weighted mean radiative
  efficiency, $\Delta M_{\rm BH}$ is the change in the mass of the BH, 
  $\Delta M_*$ is the mass of stars formed during the
  simulation, $\Delta M_W$ is the total gas mass driven beyond 10
  effective radii, and $\log M_{\rm gas}$ is the total mass remaining
  within 10 effective radii.}

\end{deluxetable*} 

\begin{figure*}
\centering
\includegraphics[width=0.7\textwidth]{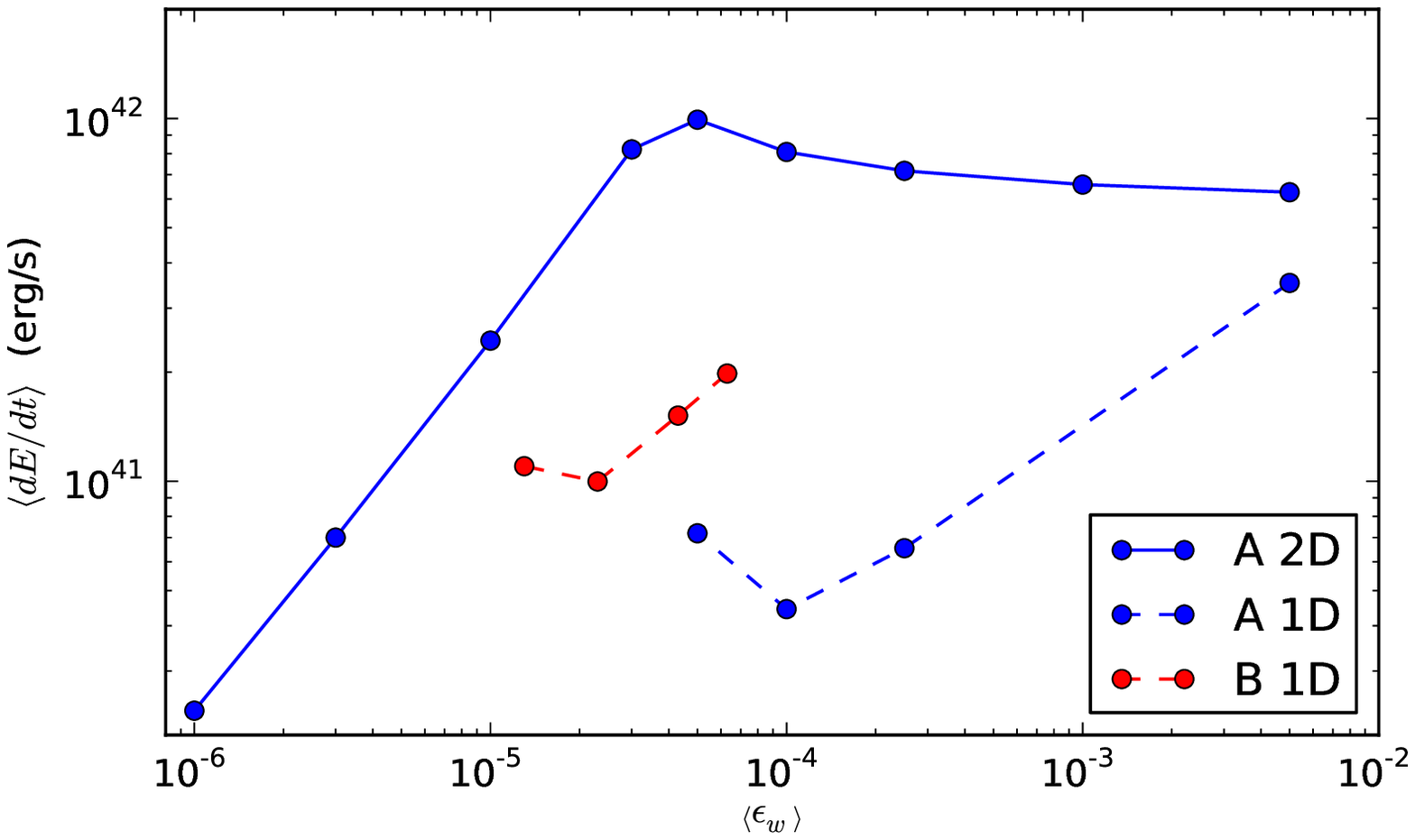}
\caption{ Time-averaged mean energy input ($\int \dot{E} \, dt / \int
  dt$) via AGN mechanical feedback vs. mass-averaged mean feedback
  efficiency ($\int \epsilon \dot{m} \, dt / \int \dot{m} \,
  dt$).  For the two-dimensional simulations, at the low-efficiency end, energy
  input falls off because the SMBH is consuming all of the mass
  available to it: $\Delta M_{\rm BH}$ is at a maximum but
  $\epsilon_W$ is still falling.  At the high-efficiency end, energy
  input approaches a constant---the simulation self-regulates to an
  energy input of $\simeq 5\times10^{41}$ erg s$^{-1}$.  This implies a
  nearly perfect, inverse relationship between $\Delta M_{\rm BH}$ and
  $\epsilon_W$ at the high-efficiency end.  The one-dimensional
  simulations have smaller energy inputs at a given efficiency,
  reflecting their smaller SMBH growth rates.}
\label{fig:efb-vs-eff}
\end{figure*}

Figure \ref{fig:efb-vs-eff} shows the mean mechanical energy input
versus the mean efficiency for one-dimensional and two-dimensional A models.  For two-dimensional
A models, the energy input is nearly constant---the SMBH accretion
self-regulates to provide energy at this rate.  The one-dimensional A models have
lower energy input rates.  That is, two-dimensional models require more energy to
reach equilibrium between inflow (due to cooling) and outflow (due to
mechanical feedback).

\section{Conclusions}
\label{sec:conclusions}
We have performed two-dimensional simulations of the entire cosmic
history (12 Gyr) of an isolated $L_*$ elliptical galaxy.  Planetary
nebulae and red giant winds produced by evolving low-mass stars serve
as the source of gas in the galaxy.  This gas finally ends up either
in the central BH, in long-lived low-mass stars (formed in the
simulation), in the ISM within the galaxy (at the end of
the simulation), or outside the galaxy as part of the intergalactic
medium.  As gas finds its way to one of those four final states, it
can engage in star formation, mass/energy injection into the ISM via
Type Ia or Type II supernovae, or radiative cooling.

The primary purpose of the code is to implement the well-developed
physical model of mechanical and radiative AGN feedback of
\citet{ciotti:09-feedback}.  This model includes momentum and energy
imparted to the gas by Compton scattering and atomic
lines via radiation from the central SMBH.  It also includes a
self-consistent model of mechanical feedback via a broad-line wind.
The primary difference between the two-dimensional code and the one-dimensional code (upon which
it is based) is that the two-dimensional code does not yet include the physics of
optically thick radiation pressure on dust grains.  Experiments with 
the one-dimensional code have shown that turning off the dust opacity does not
greatly affect the SMBH growth, although other aspects of the
simulation results such as the star formation rate are affected.  

Multi-dimensional simulations are a significant improvement over the
one-dimensional simulations because the additional degrees of freedom allow
classical instabilities such as the RT or
Kelvin--Helmholtz instabilities to operate.  The RT instability in
particular plays an important role because AGN feedback tends to
produce a hot bubble of gas near the SMBH, effectively halting
accretion.  In one dimension, the bubble cannot move away from the SMBH and
continues to suppress subsequent infall until it itself radiatively
cools.  In two dimensions, the hot bubble buoyantly moves up through the ISM and
additional cold gas generated at radii of 0.1--1 kpc can freely
fall into the center of the simulation.  

In this work, we have assumed a low specific angular momentum profile
for the model galaxy in order to match the one-dimensional simulations as closely
as possible.  The effect of angular momentum transport in galaxies
with more typical specific angular momenta is likely to be very
important.  We plan to implement angular momentum transport in future
work in order to isolate its affect on our results.

An important feature of our simulations is that we have taken care to
resolve the Bondi radius even for gas heated to the Compton
temperature by the AGN.  The gas heating terms due to AGN radiation go
as $1/r^2$, becoming much more effective as a given parcel of gas
moves toward the BH.  If the Bondi radius is resolved even for
the hot gas, then our conclusion about whether or not a given parcel
of gas makes it to the center of the simulation is secure.  If the
Bondi radius is not resolved, then radiative heating that would have
occurred between the inner radius of the simulation and the physical
Bondi radius might have increased the temperature of the gas to the
point that thermal energy dominated over gravitational energy---the
gas would not make it to the SMBH.  Thus if the Bondi radius is not
resolved for the hot as well as the cold gas, we cannot be sure
whether or not a given parcel of gas actually makes it to the SMBH.

Resolving the Bondi radius for the hot gas requires high central
spatial resolution.  The large velocities of the BAL wind entering the
grid near the center necessitate time steps of order one year.  The
timescale relevant for stellar evolution, the physical source of the
gas, is several Gyr.  We have opted to use rather coarse spatial
resolution in order to run the simulations long enough to see
evolution over the entire cosmic history of the galaxy in question.

The primary differences between the previously presented one-dimensional
simulations and the present two-dimensional simulations are that the character of
the accretion as a function of time changes from well-separated,
dramatic bursts (one dimension) to chaotic, stochastic accretion (two dimensions).  There are
two reasons for this.  First, the cold, infalling gas generated at a
few hundred parsecs tends to fragment and fall into the SMBH as several
discrete blobs (two dimensions) rather than as a coherent spherical shell (one dimension).
Second, the RT instability allows hot gas generated via radiative
heating by the AGN to move out of the way, allowing subsequent blobs
of cold gas to fall into the SMBH unimpeded.  Thus, a burst of accretion
in two dimensions is much less effective at stopping subsequent accretion than is
the case in one dimension.  

Another way of saying this is that mechanical feedback is less
effective in two dimensions than one dimension:  it takes more energy for a given galaxy
model to reach equilibrium between cooling-driven inflow and
feedback-driven outflow/heating.  At a fixed mechanical efficiency,
the SMBH self-regulates to larger mean accretion rates in order to
provide the required additional energy.  

Two-dimensional outflows are less effective than spherical ones in two
critical aspects.  less effective in protecting the SMBH from infalling
gas and less effective in ejecting gas from the galaxy.

For our models with wind efficiencies of $10^{-3}$ and $10^{-4}$,
the distribution of Eddington ratios is in reasonable agreement with
the observed fraction of galaxies above/below a given Eddington ratio
over four orders of magnitude in luminosity.  These values of the
efficiency are both physically plausible and supported by recent
observations \citep{arav:11-preprint}.

Our plans for future work include implementing optically thick
radiative transfer on dust to bring the physics of our treatment of
AGN and star-formation feedback up to the state of the art.
Furthermore, the B models for mechanical feedback are more realistic
in the sense that the efficiency and opening angle both become small
at low accretion rates, in line with observational and theoretical
expectations.  The one-dimensional simulations produced B models in good
agreement with observations \citep{ciotti:10}, but using identical
parameters for two-dimensional simulations did not result in the same good
agreement.  A parameter study of B models in two dimensions is
necessary to identity viable ranges of the parameters.  Finally, there
remain many potentially important physical processes that we have not
attempted to model in this work---energy transport via conduction and
the extent to which conduction is suppressed by tangled magnetic
fields among them.  

In this paper, we have focused on the comparison between one-dimensional and two-dimensional AGN
feedback models with nearly the same implemented physics and spatial
resolution.  A future study will put this model into better context by
comparing more thoroughly to the existing AGN feedback models in the
literature \citep[e.g.,][]{dimatteo:05, johansson:09, debuhr:10}.

\section*{Acknowledgments}
We thank Jenny Greene, Jim Stone, and Daniel Proga for
useful discussions, Nahum Arav for making results available
in advance of publication, and the anonymous referee for a very
thoughtful and helpful report.  G.S.N. was supported by the Princeton
University Council on Science and Technology and grants NASA
NNX08AH31G and NAS8-03060.  G.S.N. also made extensive use of the
computing facilities of the Princeton Institute for Computational
Science and Engineering.  J.P.O. acknowledges the support of NSF grant
AST-0707505.  L.C. was supported by the MIUR grant PRIN2008.


\end{document}